\documentclass[proceedings]{JHEP3}
\usepackage{amssymb}
\usepackage{amsfonts}
\usepackage{amsmath}
\usepackage{epsfig,multicol}

\newbox\mybox

\newcommand\fverb{\setbox\mybox=\hbox\bgroup\verb}
\newcommand\fverbdo{\egroup\medskip\noindent\fbox{\unhbox\mybox}\ }
\newcommand\fverbit{\egroup\item[\fbox{\unhbox\mybox}]}
\conference{A Q-operator for the quantum transfer matrix}
\abstract{Baxter's Q-operator for the quantum transfer matrix of the XXZ spin-chain is 
constructed employing the representation theory of quantum groups. The spectrum of this 
Q-operator is discussed and novel functional relations which describe the finite 
temperature regime of the XXZ spin-chain are derived. For non-vanishing magnetic field 
the previously known Bethe ansatz equations can be replaced by a system of quadratic 
equations which is an important advantage for numerical studies. For vanishing 
magnetic field and rational coupling values it is argued that the quantum transfer 
matrix exhibits a loop algebra symmetry closely related to the one of the classical 
six-vertex transfer matrix at roots of unity.}
\input{tcilatex}

\title{A Q-operator for the quantum transfer matrix}
\author{Christian Korff \\
Department of Mathematics, University of Glasgow, \\
University Gardens, Glasgow G12 8QW, UK\\
E-mail: c.korff@maths.gla.ac.uk}

\begin{document}

\section{Introduction}

In this work we present new identities for the description of the spectrum
of the quantum transfer matrix for the XXZ spin-chain with non-vanishing
external magnetic field $h$,%
\begin{equation}
H_{\text{XXZ}}=\frac{1}{2}\sum_{\ell =1}^{L}\left\{ \sigma _{\ell
}^{x}\sigma _{\ell +1}^{x}+\sigma _{\ell }^{y}\sigma _{\ell +1}^{y}+\Delta
(\sigma _{\ell }^{z}\sigma _{\ell +1}^{z}-1)\right\} -\frac{h}{2}\sum_{\ell
=1}^{L}\sigma ^{z}\;.  \label{HXXZ}
\end{equation}%
Here $\Delta =(q+q^{-1})/2$\ is an anisotropy parameter which is assumed to
be real. This spin-chain serves as a prototype model for other more
complicated integrable systems. It is an important toy model for the exact
computation of physical quantities such as magnetic, electric or thermal
conductivities. In this context the study of the finite temperature
behaviour of the spin-chain is of crucial importance. One method to achieve
this is the so-called quantum transfer matrix (see e.g. \cite{QTM}),%
\begin{equation}
\tau (z;w)=\limfunc{Tr}_{0}q^{\alpha ~\sigma ^{z}\otimes
1}R_{0N}(zw)R_{(N-1)0}^{t\otimes 1}(w/z)\cdots R_{02}(zw)R_{10}^{t\otimes
1}(w/z),\quad q^{\alpha }:=e^{\beta h/2}  \label{quantumT}
\end{equation}%
in terms of which the partition function of the XXZ spin-chain can be
expressed%
\begin{equation}
Z_{L}=\limfunc{Tr}_{(\mathbb{C}^{2})^{\otimes L}}e^{-\beta H_{\text{XXZ}%
}}=\lim_{N\rightarrow \infty }\limfunc{Tr}_{(\mathbb{C}^{2})^{\otimes
N}}\tau (z=1;w=e^{-\beta ^{\prime }/N})^{L},\qquad \beta ^{\prime }=\beta
(q-q^{-1})\ .  \label{Z}
\end{equation}%
Let us explain the various objects appearing in the definition. The variable 
$\beta >0$ denotes the inverse temperature of the system and for convenience
we have introduced a \textquotedblleft twist angle\textquotedblright\ $%
\alpha =\beta h/2\ln q$ with $h$ being the magnetic field in (\ref{HXXZ}).
The quantum transfer matrix is built out of the well-known six-vertex $R$%
-matrix which acts on the tensor product $\mathbb{C}^{2}\otimes \mathbb{C}%
^{2},$%
\begin{equation}
R=\left( 
\begin{array}{cc}
\;a\; & 0 \\ 
0 & \;b\;%
\end{array}%
\right) \otimes \sigma ^{+}\sigma ^{-}+\left( 
\begin{array}{cc}
\;b\; & 0 \\ 
0 & \;a\;%
\end{array}%
\right) \otimes \sigma ^{-}\sigma ^{+}+c\,\sigma ^{+}\otimes \sigma
^{-}+c^{\prime }\,\sigma ^{-}\otimes \sigma ^{+}\ .  \label{R}
\end{equation}%
Here the parametrization of the Boltzmann weights $a,b,c,c^{\prime }$ is
chosen as follows\footnote{%
Below we will compare the results in this paper with the known properties
about the quantum transfer matrix as they can be found in e.g. \cite{Klue04}%
. To this end it is helpful to identify the definition of the Boltzmann
weights in \cite{Klue04} on page 11, equation (2) with ours by setting $%
z=e^{i\gamma w}$ and $q=e^{i\gamma }$. Note that on page 16 in \cite{Klue04}%
\ a rotation in the complex plane is performed replacing $w\rightarrow iv$;
see equations (23) and (24) therein.}%
\begin{equation}
a=1,\quad b=\frac{(1-z)q}{1-zq^{2}},\quad c=\frac{1-q^{2}}{1-zq^{2}},\quad
c^{\prime }=c\,z\ .  \label{abc}
\end{equation}%
The upper index $t\otimes 1$ in (\ref{quantumT}) stands for transposition in
the first factor. We recall that the six-vertex $R$-matrix gives rise to a
classical statistical mechanics system which is described by the \emph{%
classical} six-vertex transfer matrix (as opposed to \emph{quantum}),%
\begin{equation}
t_{\text{6v}}(z)=\limfunc{Tr}_{0}R_{0L}(z)\cdots R_{01}(z)\;.  \label{classT}
\end{equation}%
This classical physical system is connected with the above quantum
spin-chain through the relation (we set temporarily $h=0$)%
\begin{equation}
H_{\text{XXZ}}=(q-q^{-1})~\left. z\frac{d}{dz}\ln t_{\text{6v}%
}(z)\right\vert _{z=1}\ .  \label{Hfromt}
\end{equation}%
The prefactor in the last equation explains the introduction of the
renormalised temperature variable $\beta ^{\prime }$ in (\ref{Z}). As an
immediate consequence of (\ref{Hfromt}) one has the following identity for
the density matrix%
\begin{equation}
\lim_{N\rightarrow \infty }\left( t_{\text{6v}}(1)^{-1}t_{\text{6v}%
}(e^{-\beta ^{\prime }/N})\right) ^{N}=e^{-\beta H_{\text{XXZ}}}\ .
\label{density}
\end{equation}%
This rewriting is inspired by the Trotter formula in the path-integral
formalism of quantum field theory and $N$ is referred to as Trotter number 
\cite{Trotter59}. The last expression (\ref{density}) can be conveniently
expressed in terms of the quantum transfer matrix (\ref{quantumT}); see the
review \cite{Klue04} and references therein for details. Note in particular
that for this construction to work the Trotter number $N$ has to be even and
we shall work throughout this paper with the convention%
\begin{equation}
N=2n\ .  \label{Nn}
\end{equation}

In the thermodynamic limit when the physical system size $L$ tends to
infinity the partition function (\ref{Z}) can be approximated by the largest
eigenvalue $\Lambda _{N}$ of the quantum transfer matrix (\ref{quantumT}),%
\begin{equation}
L\gg 1:\;Z_{L}=\lim_{N\rightarrow \infty }\limfunc{Tr}_{(\mathbb{C}%
^{2})^{\otimes N}}\tau (1;e^{-\beta ^{\prime }/N})^{L}\approx
\lim_{N\rightarrow \infty }(\Lambda _{N})^{L}\ .  \label{approxZ}
\end{equation}%
Thus, the introduction of the quantum transfer matrix (\ref{quantumT})
allows one to efficiently investigate the finite temperature regime. Instead
of having to compute multiple excited states and energies of the quantum
spin-chain (\ref{HXXZ}) at finite temperature, one only needs to compute a
single eigenvalue and eigenstate of the quantum transfer matrix (\ref%
{quantumT}). Similar simplifications occur also in the computation of finite
temperature correlation functions; see for instance \cite{GKS04}.\smallskip

There are some technical complications, however. We refer the reader for the
following statements to \cite{Klue04} and references therein. The algebraic
properties of the quantum transfer matrix resemble closely the ones of the
classical six-vertex transfer matrix (\ref{classT}) and as a result one can
compute the eigenstates and eigenvalues of the quantum transfer matrix by
similar methods as in the classical case, i.e. the algebraic Bethe ansatz.
Via this route one obtains expressions for the eigenvalues of (\ref{quantumT}%
). In particular the largest eigenvalue $\Lambda _{N}$ can be cast into the
form%
\begin{equation}
\Lambda _{N}(u)=\frac{e^{\frac{\beta h}{2}}\phi (u-i)\mathcal{Q}(u+2i)+e^{-%
\frac{\beta h}{2}}\phi (u+i)\mathcal{Q}(u-2i)}{\mathcal{Q}(u)[\sinh \frac{%
\gamma }{2}(u-2i+i\tau )\sinh \frac{\gamma }{2}(u+2i-i\tau )]^{\frac{N}{2}}}%
~,  \label{Lam}
\end{equation}%
where we have set $z=e^{\gamma u},\;w=e^{-\beta ^{\prime }/N}=e^{-i\gamma
\tau },\;q=e^{i\gamma }$ and introduced the functions%
\begin{equation}
\phi (u)=[\sinh \tfrac{\gamma }{2}(u-i+i\tau )\sinh \tfrac{\gamma }{2}%
(u-2i+i\tau )]^{n},\qquad \mathcal{Q}(u)=\tprod_{j=1}^{n}\sinh \tfrac{\gamma 
}{2}(u-u_{j})\;.  \label{Qtrig}
\end{equation}%
The quantities $u_{j}$ are solutions of the following system of non-linear
equations%
\begin{equation}
\frac{\phi (u_{j}+i)}{\phi (u_{j}-i)}=-e^{\beta h}\frac{\mathcal{Q}(u_{j}+2i)%
}{\mathcal{Q}(u_{j}-2i)},\qquad j=1,...,n\ .  \label{BAE}
\end{equation}%
Compare with equations (25-30) in \cite{Klue04}. The analytic solutions to
the last set of equations, known as Bethe ansatz equations, are not known.
Furthermore, in the Trotter limit $N\rightarrow \infty $ the distribution of
the Bethe roots $u_{j}$ remains discrete and cannot be approximated by
continuous density functions as it is the case for the classical transfer
matrix (\ref{classT}). Instead one has to rely on the numerical solution of
a non-linear integral equation. The derivation of this integral equation as
well as other properties of the quantum transfer matrix are based on
numerical studies of the above Bethe ansatz equations (\ref{BAE}).\smallskip

With this in mind it is worthwhile to study alternative methods of deriving
the spectrum, and in particular the largest eigenvalue, which lead to
systems of equations simpler than the Bethe ansatz equations (\ref{BAE}).
The purpose of this article is to diagonalise the quantum transfer matrix
employing Baxter's idea of an auxiliary matrix known as Q-operator \cite%
{Bx82}. The eigenvalues of the Q-operator give the $Q$-function in (\ref%
{Qtrig}) and we will derive a set of functional relations analogous to the
ones first obtained in the context of conformal field theory \cite{BLZ97}.
The case of the finite XXZ spin-chain has been investigated in \cite{RW02}
and \cite{CKQ3,CKQ4,CKQ7}. In particular, we will follow closely the
treatment for the twisted XXZ spin-chain given in \cite{CKQ7}.\smallskip

In order to construct the $Q$-operator we will use the representation theory
of quantum groups. The main results for $h\neq 0$ in (\ref{HXXZ}) are novel
identities for the spectrum of the quantum transfer matrix and a simpler set
of equations which imply the Bethe ansatz equations (\ref{BAE}) but are 
\emph{quadratic} instead of order $N$. One of the main results in this paper
is that the largest eigenvalue $\Lambda _{N}$ of (\ref{quantumT}) can be
expressed in terms of one polynomial%
\begin{equation}
Q^{+}(z)=\sum_{k=0}^{n}e_{k}^{+}(-z)^{k},  \label{Qpm1}
\end{equation}%
whose coefficients $e_{k}^{+}$ (with $e_{0}^{+}=1$) solve the following
system of \emph{quadratic} equations%
\begin{equation}
e_{n}^{+}\sum_{k+l=m}\binom{n}{k}\binom{n}{l}(wq)^{k-l}=\sum_{k+l=m}\frac{%
\sinh [\frac{\beta h}{2}-i\gamma (k-l)]}{\sinh [\beta h/2]}%
~e_{k}^{+}e_{n-l}^{+}\ .  \label{result}
\end{equation}%
Here the summation convention in (\ref{result}) is to be understood as
follows. First fix the variable $m$ in the allowed range $m=1,...,N-1$ and
then sum in (\ref{result}) over all possible values for $k,l$ such that $%
k+l=m$. Thus, one obtains in total $N-1$ coupled quadratic equations. For
real $z$ the largest eigenvalue of the quantum transfer matrix is then given
by 
\begin{equation}
\Lambda _{N}(z;w)=\frac{e^{\beta h}Q^{+}(zq^{-2})Q^{-}(zq^{2})-e^{-\beta
h}Q^{+}(zq^{2})Q^{-}(zq^{-2})}{\sinh [\beta h/2][(zwq-1)(z/wq-1)]^{N/2}}~,
\label{LamN}
\end{equation}%
where $Q^{-}$ is the reciprocal polynomial of $Q^{+}$, i.e.%
\begin{equation*}
Q^{-}(z)=\sum_{k=0}^{n}\frac{e_{n-k}^{+}}{e_{n}^{+}}%
(-z)^{k}=z^{n}Q^{+}(z^{-1})/e_{n}^{+}\ .
\end{equation*}%
In the critical regime, $q\in \mathbb{S}^{1}$, the coefficients $e_{k}^{+}$
obey the additional constraint%
\begin{equation}
(e_{k}^{+})^{\ast }=e_{n-k}^{+}/e_{n}^{+}\ .  \label{cce}
\end{equation}%
There are in general many solutions to the equations (\ref{result}) (similar
as there are multiple solutions to the Bethe ansatz equations) describing a
subset of the spectrum of (\ref{quantumT}). The largest eigenvalue of the
quantum transfer matrix $\Lambda _{N}$ appears to be always among them; this
has been numerically verified for $N=2,4,6,8,10,12,14,16~\footnote{%
It appears that there is a difference between $n=N/2$ odd and even; compare
with the footnote in \cite{GKS04}, Section 2.6. For $n$ odd the largest
eigenvalue might have less than $n$ Bethe roots. Nevertheless, in the cases $%
N=2,6,10$ we have numercially verified that (\ref{result}) and (\ref{LamN})
still hold true.}$ and for all of these cases the total number of solutions
to (\ref{result}) is found to be $2^{n}$.

The identities (\ref{result}) and (\ref{LamN}) are special cases of more
generally valid expressions for the spectrum of the quantum transfer matrix;
see equations (\ref{quantumW}) and (\ref{TQpm}) in the text. We discuss
these identities in the context of higher spin quantum transfer matrices and
the associated fusion hierarchy. This is motivated by the definition of a
trace functional which might be relevant for the computation of correlation
functions. Underlying the derivation of the main result is a concrete
operator construction for $Q^{\pm }$ which is important to extract further
information on the spectrum and also applies to the case of vanishing
magnetic field, $h=0$ in (\ref{HXXZ}), albeit one has then to restrict the
deformation parameter $q$ to a root of unity, $q^{\ell }=1,~\ell >2$. In
this case, we argue that the quantum transfer matrix exhibits a loop algebra
symmetry\ just as the classical six-vertex transfer matrix \cite{DFM} albeit
in a different representation.

The outline of the article is as follows. In Section 2 we connect the
structure of the quantum transfer matrix to the representation theory of the
quantum affine algebra $U_{q}(\widehat{sl}_{2})$. In particular, we discuss
a specific dual representation and the corresponding $L$-operator. This will
yield the commutation relations between the quantum monodromy matrix
elements corresponding to (\ref{quantumT}) and the Chevalley-Serre
generators of $U_{q}(\widehat{sl}_{2})$ which allows us to prove the loop
algebra symmetry $U(s\tilde{l}_{2})$ of the quantum transfer matrix at roots
of unity and for vanishing magnetic field. As a preparatory step for the
discussion of the $Q$-operator we introduce the quantum fusion hierarchy,
i.e. the higher spin analogues of the quantum transfer matrix.

In Section 3 we construct the $Q$-operator for the quantum transfer matrix
and discuss its properties and functional relations. The proofs can be found
in appendices A and B. For the construction of the Q-operator one has
carefully to distinguish between the case of generic $q$ and $q$ a root of
unity. In the former case the auxiliary space of the $Q$-operator is
infinite-dimensional and one needs to introduce a boundary parameter (the
external magnetic field $h$) in order to ensure convergence. At a root of
unity the auxiliary space is finite-dimensional and the construction of $Q$
then also applies to the case $h=0$. However, some of the functional
relations which hold true at $h\neq 0$ then cease to be valid.

Section 4 is devoted to a special Q-operator functional equation, the
Wronskian relation, which suffices to compute the spectrum of the quantum
transfer matrix and implies the Bethe ansatz equations. At the end we
discuss some special solutions which contain the largest eigenvalue and are
based on numerical evidence.

Section 5 contains the conclusions.

\section{The Quantum transfer matrix and representation theory}

As a preparatory step for the construction of the $Q$-operator for the
quantum transfer matrix let us first analyse the definition (\ref{quantumT})
from a representation theoretic point of view. This will enable us to define
the corresponding $L$-operator from which the $Q$-operator will be built.

Recall that the six-vertex $R$-matrix is an intertwiner of the quantum
affine algebra $U_{q}(\widehat{sl}_{2})$ with respect to the tensor product
of the two-dimensional evaluation representation. The quantum affine algebra 
$U_{q}(\widehat{sl}_{2})$ is generated from the Chevalley-Serre elements
subject to the relations%
\begin{equation}
q^{h_{i}}q^{h_{j}}=q^{h_{j}}q^{h_{i}},\quad
q^{h_{i}}e_{j}q^{-h_{i}}=q^{A_{ij}}e_{j},\quad
q^{h_{i}}f_{j}q^{-h_{i}}=q^{-A_{ij}}f_{j},\quad \lbrack e_{i},f_{j}]=\delta
_{ij}\frac{q^{h_{i}}-q^{-h_{i}}}{q-q^{-1}}
\end{equation}%
and%
\begin{equation}
x_{i}^{3}x_{j}-[3]_{q}x_{i}^{2}x_{j}x_{i}+[3]_{q}x_{i}x_{j}x_{i}^{2}-x_{j}x_{i}^{3}=0,\quad x=e,f\ .
\end{equation}%
Here the indices $i,j$ take the values 0,1 and $A_{ij}$ is the Cartan matrix
of $\widehat{sl}_{2}$. The evaluation homomorphism $ev_{z}:U_{q}(\widehat{sl}%
_{2})\rightarrow U_{q}(sl_{2})$ defined by%
\begin{equation}
ev_{z}(e_{0})=z~f,\quad ev(f_{0})=z^{-1}e,\quad ev(q^{h_{0}})=q^{-h}
\end{equation}%
and%
\begin{equation}
ev(e_{1})=e,\quad ev(f_{1})=f,\quad ev(q^{h_{1}})=q^{h}\ .
\end{equation}%
An evaluation representation is now obtained by combining the evaluation
homomorphism with any finite-dimensional representation of $U_{q}(sl_{2})$,
in particular we can choose the two-dimensional, spin 1/2 representation in
terms of Pauli matrices,%
\begin{equation}
\pi (e)=\sigma ^{+},\quad \pi (f)=\sigma ^{-},\quad \pi (q^{h})=q^{\sigma
^{z}}\ .  \label{spin1/2}
\end{equation}%
The six-vertex R-matrix then intertwines the tensor product representation $%
\pi _{z}\otimes \pi _{1}$ with $\pi _{z}=\pi \circ ev_{z}$.

Given any representation $\rho :U_{q}(\widehat{sl}_{2})\rightarrow \limfunc{%
End}V$ over some finite-dimensional vector space $V$ we define the following
representation over its dual space $V^{\ast }$,%
\begin{equation}
\rho ^{\ast }:U_{q}(\widehat{sl}_{2})\rightarrow \limfunc{End}V^{\ast
},\qquad \left\langle \rho ^{\ast }(x)v^{\ast },w\right\rangle
:=\left\langle v^{\ast },\rho (\gamma ^{-1}(x))w\right\rangle ,\quad x\in
U_{q}(\widehat{sl}_{2}),\;w\in V\ .  \label{star}
\end{equation}%
Here $\gamma $ is the antipode which is defined on the Chevalley-Serre
generators as follows,%
\begin{eqnarray}
\gamma (e_{i}) &=&-q^{-h_{i}}e_{i},\qquad \gamma
(f_{i})=-f_{i}q^{h_{i}},\qquad \gamma (q^{\pm h_{i}})=q^{\mp h_{i}}
\label{antip} \\
\gamma ^{-1}(e_{i}) &=&-e_{i}q^{-h_{i}},\qquad \gamma
^{-1}(f_{i})=-q^{h_{i}}f_{i},\qquad \gamma ^{-1}(q^{\pm h_{i}})=q^{\mp
h_{i}}\ .
\end{eqnarray}%
If we canonically identify $V^{\ast }$ with $V$ the representation $\rho
^{\ast }$ in terms of matrices is simply given by%
\begin{equation}
\rho ^{\ast }(x)=(\rho (\gamma ^{-1}(x)))^{t},\qquad x\in U_{q}(\widehat{sl}%
_{2})
\end{equation}%
with $t$ denoting the transpose. Note that the definition (\ref{antip}) is
compatible with the following choice for the coproduct%
\begin{equation}
\Delta (e_{i})=e_{i}\otimes 1+q^{h_{i}}\otimes e_{i},\qquad \Delta
(f_{i})=f_{i}\otimes q^{-h_{i}}+1\otimes f_{i},\qquad \Delta
(q^{h_{i}})=q^{h_{i}}\otimes q^{h_{i}}\ .  \label{cop}
\end{equation}%
Setting $\rho =\pi $, the two-dimensional evaluation module, we are
interested in finding the intertwiner%
\begin{equation}
L^{\ast }~\Delta ^{\ast }(x)=\Delta _{\text{op}}^{\ast }(x)L^{\ast }\quad 
\text{with}\quad \Delta ^{\ast }=(1\otimes \pi ^{\ast })\Delta \quad \text{%
and}\quad L^{\ast }\in U_{q}(sl_{2})\otimes \limfunc{End}V^{\ast }\ .
\label{Linter}
\end{equation}%
Explicitly, the coproduct relations for the Chevalley-Serre generators read%
\begin{eqnarray}
\Delta ^{\ast }(e_{1}) &=&e_{1}\otimes 1-q^{1+h_{1}}\otimes \sigma
^{-},\quad \quad \Delta _{\text{op}}^{\ast }(e_{1})=e_{1}\otimes q^{-\sigma
^{z}}-q~1\otimes \sigma ^{-}~, \\
\Delta ^{\ast }(f_{1}) &=&f_{1}\otimes q^{\sigma ^{z}}-q^{-1}1\otimes \sigma
^{+},\quad \quad \Delta _{\text{op}}^{\ast }(f_{1})=f_{1}\otimes
1-q^{-1-h_{1}}\otimes \sigma ^{+}~,
\end{eqnarray}%
and%
\begin{eqnarray}
\pi ^{\ast }(e_{1}) &=&-(\sigma ^{+}q^{-\sigma ^{z}})^{t}=-q~\sigma
^{-},\quad \quad \pi ^{\ast }(f_{1})=-(q^{\sigma ^{z}}\sigma
^{-})^{t}=-q^{-1}\sigma ^{+},  \label{spin1/2*} \\
\pi ^{\ast }(e_{0}) &=&-(\sigma ^{-}q^{\sigma ^{z}})^{t}=-q~\sigma
^{+},\quad \quad \pi ^{\ast }(f_{0})=-(q^{-\sigma ^{z}}\sigma
^{+})^{t}=-q^{-1}\sigma ^{-}~.
\end{eqnarray}%
We will see below that the intertwiner (\ref{Linter}) specializes to the $R$%
-matrix used in the definition of the quantum transfer matrix (\ref{quantumT}%
). Solving the above intertwining condition (\ref{Linter}) for an evaluation
module with central charge zero, i.e. $q^{h_{0}}=q^{-h_{1}},$ we find%
\begin{equation}
L^{\ast }(z)=\left( 
\begin{array}{cc}
zq^{-\frac{h_{1}+1}{2}}-q^{\frac{h_{1}+1}{2}} & (q^{-1}-q)e_{1}q^{-\frac{%
h_{1}+1}{2}} \\ 
z(q^{-1}-q)q^{\frac{h_{1}+1}{2}}f_{1} & zq^{\frac{h_{1}-1}{2}}-q^{-\frac{%
h_{1}-1}{2}}%
\end{array}%
\right) \ .  \label{L1}
\end{equation}%
For comparison and in order to keep this article self-contained recall that
the conventional $L$-operator reads%
\begin{equation}
L(z)=\left( 
\begin{array}{cc}
zq^{\frac{h_{1}+1}{2}}-q^{-\frac{h_{1}+1}{2}} & z(q-q^{-1})q^{\frac{h_{1}+1}{%
2}}f_{1} \\ 
(q-q^{-1})e_{1}q^{-\frac{h_{1}+1}{2}} & zq^{-\frac{h_{1}-1}{2}}-q^{\frac{%
h_{1}-1}{2}}%
\end{array}%
\right) \ .
\end{equation}%
In light of the following identity for the universal R-matrix $(1\otimes
\gamma ^{-1})\boldsymbol{R}=\boldsymbol{R}^{-1}$ it would be more natural to
use $-z^{-1}L^{\ast }(z)$ as intertwiner for the dual representation $\pi
^{\ast }$. However, we wish for later purposes to keep the $L^{\ast }$%
-operator polynomial in $z$ instead of $z^{-1}$. Evaluating this intertwiner
in the two-dimensional spin 1/2 representation (\ref{spin1/2}) yields%
\begin{equation}
R^{\pi ,\pi ^{\ast }}(z)=\frac{(\pi \otimes 1)L^{\ast }(z)}{zq^{-1}-q}%
=\left( R(z)^{-1}\right) ^{1\otimes t}=[R_{21}(z^{-1})]^{1\otimes t}\ ,
\end{equation}%
which is one of the R-matrices used in the definition of the quantum
transfer matrix (\ref{quantumT}). Thus, we conclude that one lattice row
associated with the monodromy matrix of the quantum transfer matrix
corresponds to the following quantum group module 
\begin{equation}
\pi _{z}\otimes \mathfrak{M}_{w}^{(N)},\;\quad \quad \mathfrak{M}_{w}^{(N)}=~%
\underset{N}{\underbrace{\pi _{w}^{\ast }\otimes \pi _{w^{-1}}\cdots \otimes
\,\pi _{w}^{\ast }\otimes \pi _{w^{-1}}}}\;.  \label{module}
\end{equation}%
From this we immediately deduce that the quantum transfer matrix $\tau (z;w)$
block decomposes with respect to the following alternating spin operator,%
\begin{equation}
\lbrack \tau (z;w),S_{A}]=0,\qquad S_{A}:=\frac{1}{2}\sum_{k=1}^{N}(-)^{k}%
\sigma _{k}^{z}\ ,  \label{SA}
\end{equation}%
since the quantum monodromy matrix%
\begin{equation}
R_{0N}(zw)R_{(N-1)0}^{t\otimes 1}(w/z)\cdots R_{02}(zw)R_{10}^{t\otimes
1}(w/z)=\left( 
\begin{array}{cc}
A(z) & B(z) \\ 
C(z) & D(z)%
\end{array}%
\right)   \label{mom}
\end{equation}%
is an intertwiner with respect to the tensor product (\ref{module}). More
generally, we have%
\begin{equation}
\lbrack
A,q^{H_{1}}]=[D,q^{H_{1}}]=0,\;q^{H_{1}}Bq^{-H_{1}}=q^{-2}B,%
\;q^{H_{1}}Cq^{-H_{1}}=q^{2}C,\qquad q^{H_{1}}=q^{2S_{A}}\ .
\end{equation}%
Denoting the Chevalley-Serre generators acting in quantum space (\ref{module}%
) by capital letters, $\{E_{1},E_{0},F_{1},F_{0},H_{0}=-H_{1}\},$ one finds
the analogous commutation relations as in the classical case; see equations
(14-15) in \cite{CKtwist}. The difference between the relations for the
classical and the quantum transfer matrix is purely in the explicit form of
the quantum group generators which is fixed through the identification of
the quantum group module $\mathfrak{M}_{w}^{(N)}$in (\ref{module}). For
instance, employing (\ref{cop}) and (\ref{spin1/2}), (\ref{spin1/2*}) one has%
\begin{eqnarray}
E_{1} &=&\sum_{k=1}^{N}\varepsilon _{k}q^{\frac{1-\varepsilon _{k}}{2}%
}\left( \tprod_{j<k}q^{\varepsilon _{j}\sigma _{j}^{z}}\right) \sigma
_{k}^{\varepsilon _{k}},\qquad \varepsilon _{k}=(-1)^{k} \\
F_{1} &=&\sum_{k=1}^{N}\varepsilon _{k}q^{\frac{\varepsilon _{k}-1}{2}%
}\sigma _{k}^{-\varepsilon _{k}}\left( \tprod_{j>k}q^{-\varepsilon
_{j}\sigma _{j}^{z}}\right) ,
\end{eqnarray}%
and so forth. Following the same line of argument as presented in \cite%
{CKtwist} (c.f. equations (23-24) therein) one then proves that for zero
magnetic field $h=0$ and $q$ being a primitive root of unity of order $\ell $
the quantum transfer matrix (\ref{quantumT}) enjoys a loop algebra symmetry $%
U(\widetilde{sl}_{2})$ in the sectors%
\begin{equation}
2S_{A}=0\func{mod}\ell \ .  \label{comensurate}
\end{equation}%
The Chevalley-Serre generators of the loop algebra $U(\widetilde{sl}_{2})$
are obtained from the restricted quantum group (compare with the discussion
in \cite{CP}) via taking the following limit from generic $q^{\prime }$ to
the root of unity value $q$,%
\begin{equation*}
q^{\ell }=1:\qquad E_{1}^{(\ell ^{\prime })}:=\lim_{q^{\prime }\rightarrow
q}E_{1}^{\ell ^{\prime }}/[\ell ^{\prime }]_{q^{\prime }}!,\qquad \lbrack
x]_{q}:=\frac{q^{x}-q^{-x}}{q-q^{-1}},\quad \ell ^{\prime }=\left\{ 
\begin{array}{cc}
\ell , & \text{if }\ell \text{ is odd} \\ 
\ell /2, & \text{if }\ell \text{ is even}%
\end{array}%
\right. \ .
\end{equation*}%
Analogous expressions hold for the remaining generators. All of the
Chevalley-Serre generators, $\{E_{1}^{(\ell ^{\prime })},E_{0}^{(\ell
^{\prime })},F_{1}^{(\ell ^{\prime })},F_{0}^{(\ell ^{\prime })}\}$, commute
in the commensurate sectors (\ref{comensurate}) with the quantum transfer
matrix, e.g.%
\begin{equation}
\lbrack \tau (z;w),E_{1}^{(\ell ^{\prime })}]=0\qquad \text{for}\;\
h=0,\quad \;q^{\ell }=1\quad \;\text{and}\quad \;2S_{A}=0\func{mod}\ell \;.
\label{loop}
\end{equation}%
This result for the quantum transfer matrix is analogous to the loop
symmetry of the classical transfer matrix first discovered in \cite{DFM},
albeit via a different proof.

\subsection{Transformation under spin reversal}

In order to show the existence of two independent solutions to the
TQ-equation, we discuss the behaviour of the quantum transfer matrix under
spin-reversal. From the elementary identities%
\begin{eqnarray*}
(1\otimes \sigma ^{x})R(z)(1\otimes \sigma ^{x}) &=&(\sigma ^{x}\otimes z^{-%
\frac{\sigma ^{z}}{2}})R(z)(\sigma ^{x}\otimes z^{\frac{\sigma ^{z}}{2}}) \\
(1\otimes \sigma ^{x})R^{\ast }(z)(1\otimes \sigma ^{x}) &=&(\sigma
^{x}\otimes z^{\frac{\sigma ^{z}}{2}})R^{\ast }(z)(\sigma ^{x}\otimes z^{-%
\frac{\sigma ^{z}}{2}})
\end{eqnarray*}%
we infer that the quantum transfer matrix (\ref{quantumT}) transforms as%
\begin{equation}
w^{S^{z}}\mathfrak{R~}\tau _{\alpha }(z;w)~w^{S^{z}}\mathfrak{R}=\tau
_{-\alpha }(z;w)  \label{RtauR}
\end{equation}%
under the involution $w^{S^{z}}\mathfrak{R}$ with%
\begin{equation}
\mathfrak{R}=\tprod_{j=1}^{N}\sigma _{j}^{x}\qquad \text{and}\qquad S^{z}=%
\frac{1}{2}\sum_{j=1}^{N}\sigma _{j}^{z}\;.  \label{RS}
\end{equation}%
In addition, the quantum transfer matrix obeys another identity. First we
observe that the following equations for the Boltzmann weights (\ref{abc})
hold true,%
\begin{equation}
a_{z^{-1}}=1,\quad b_{z^{-1}}=1/b_{zq^{-2}},\quad
c_{z^{-1}}=-q~c_{zq^{-2}}^{\prime }/b_{zq^{-2}},\quad c_{z^{-1}}^{\prime
}=-q^{-1}c_{zq^{-2}}/b_{zq^{-2}}\ .
\end{equation}%
Employing the identities%
\begin{eqnarray*}
R(z^{-1}) &=&\frac{1}{b_{zq^{-2}}}~(\sigma ^{x}\otimes (-q)^{-\frac{\sigma
^{z}}{2}})R(zq^{-2})^{1\otimes t}(\sigma ^{x}\otimes (-q)^{\frac{\sigma ^{z}%
}{2}}) \\
R^{\ast }(z^{-1}) &=&\frac{1}{b_{z^{-1}q^{-2}}}~(\sigma ^{x}\otimes (-q)^{%
\frac{\sigma ^{z}}{2}})R^{\ast }(zq^{2})^{1\otimes t}(\sigma ^{x}\otimes
(-q)^{-\frac{\sigma ^{z}}{2}})
\end{eqnarray*}%
we then easily find the expression for the transpose of the quantum transfer
matrix,%
\begin{eqnarray}
\tau _{-\alpha }(z^{-1},w^{-1}) &=&\limfunc{Tr}_{0}q^{-\alpha ~\sigma
^{z}\otimes 1}R_{0N}(z^{-1}w^{-1})R_{0N-1}^{\ast }(w/z)\cdots
R_{02}(z^{-1}w^{-1})R_{01}^{\ast }(w/z)  \notag \\
&=&\frac{\tau _{\alpha }(z,wq^{-2})^{t}}{b_{zwq^{-2}}^{n}b_{wq^{-2}/z}^{n}},
\label{tautransp}
\end{eqnarray}%
where we have used that $[S_{A},\tau _{\alpha }(z,w)]=0$.

\subsection{The quantum fusion hierarchy}

The quantum transfer matrix (\ref{quantumT}) commutes with an infinite
family of higher spin transfer matrices. In close analogy with the classical
six-vertex model we define for $d\in \mathbb{N}$ the following family of
transfer matrices,\footnote{%
In the following we shall often suppress the explicit dependence on the
temperature parameter $w$.}%
\begin{equation}
\tau ^{(d-1)}(z;w)=\limfunc{Tr}_{\pi ^{(d-1)}}q^{\alpha h\otimes
1}L_{N}(zw)L_{N-1}^{\ast }(z/w)\cdots L_{2}(zw)L_{1}^{\ast }(z/w),\qquad
q^{\alpha }=e^{\frac{h\beta }{2}}\ ,  \label{fusedt}
\end{equation}%
where the representation in auxiliary space has been replaced by the spin $%
(d-1)/2$ evaluation module $\pi ^{(d-1)},$%
\begin{equation}
\pi ^{(d-1)}(e)\left\vert k\right\rangle =[d-k]_{q}\left\vert
k-1\right\rangle ,\;\pi ^{(d-1)}(f)\left\vert k\right\rangle
=[k+1]_{q}\left\vert k+1\right\rangle ,\;\pi ^{(d-1)}(q^{h})\left\vert
k\right\rangle =q^{d-2k-1}\left\vert k\right\rangle ,
\end{equation}%
where the index $k$ labelling the basis vectors of the representation takes
values in the set $k=0,1,2,...,d-1$ and we set $e\left\vert 0\right\rangle
=f\left\vert d-1\right\rangle =0.$ Our motivation for introducing (\ref%
{fusedt}) is twofold. They are natural objects to consider from a
representation theoretic point of view and we will encounter them when
deriving functional relations for the $Q$-operator in the subsequent
section. The other reason is their extension to complex dimension $d\in 
\mathbb{C}$ which is closely related to the trace functional used in recent
formulations for correlation functions \cite{BJMST}.

Setting $d=2$ we identify $\pi ^{(1)}\equiv \pi $ in (\ref{spin1/2}) and
recover the quantum transfer matrix via the relation 
\begin{equation}
\tau (z;w)=\frac{(-z/w)^{-n}\tau ^{(1)}(z;w)}{%
(zwq-q^{-1})^{n}(wq/z-q^{-1})^{n}}=\frac{\tau ^{(1)}(z;w)}{%
(zwq-q^{-1})^{n}(z/wq-q)^{n}}\ .  \label{tauid}
\end{equation}%
The spin $0$ representation yields the quantum determinant,%
\begin{eqnarray}
\tau ^{(0)}(z;w) &=&(zwq^{\frac{1}{2}}-q^{-\frac{1}{2}})^{n}(zq^{-\frac{1}{2}%
}/w-q^{\frac{1}{2}})^{n}  \notag \\
&=&(zw-q^{-1})^{n}(z/w-q)^{n}=(zwq-1)^{n}(z/wq-1)^{n}  \label{quantumdet}
\end{eqnarray}%
Similar to the classical six-vertex model, the higher-spin quantum transfer
matrices $\tau ^{(d)}$ satisfy a functional relation known as the fusion
hierarchy,%
\begin{equation}
\tau ^{(d-1)}(zq^{d})\tau ^{(1)}(z)=\tau ^{(0)}(zq^{-1})\tau
^{(d-2)}(zq^{d+1})+\tau ^{(0)}(zq)\tau ^{(d)}(zq^{d-1})  \label{quantumFus}
\end{equation}%
which is a corollary of the decomposition of the tensor product $\pi
_{zq^{d}}^{(d-1)}\otimes \pi _{z}^{(1)}$ described by the exact sequence%
\begin{equation}
0\rightarrow \pi _{zq^{d+1}}^{(d-2)}\overset{\imath }{\hookrightarrow }\pi
_{zq^{d}}^{(d-1)}\otimes \pi _{z}^{(1)}\overset{p}{\rightarrow }\pi
_{zq^{d-1}}^{(d)}\rightarrow 0\ .  \label{seq1}
\end{equation}%
Since the auxiliary spaces in the quantum transfer matrices are the same as
in the classical case, we can use the same representation theoretic results
to derive all relevant functional relations. What changes in the transition
from \textquotedblleft classical\textquotedblright\ to \textquotedblleft
quantum\textquotedblright\ are the coefficient functions which appear in the
respective functional equation. For instance, the coefficients $\tau
^{(0)}(zq^{-1}),\ \tau ^{(0)}(zq)$ follow from the identities%
\begin{eqnarray}
(\pi ^{(d-1)}\otimes 1)L_{13}(zq^{d})R_{23}(z)(\imath \otimes 1)
&=&(z-1)(\imath \otimes 1)(\pi ^{(d-2)}\otimes 1)L(zq^{d+1})  \notag \\
(p\otimes 1)(\pi ^{(d-1)}\otimes 1)L_{13}(zq^{d})R_{23}(z) &=&\left(
zq-q^{-1}\right) (\pi ^{(d)}\otimes 1)L(zq^{d-1})(p\otimes 1)
\end{eqnarray}%
and%
\begin{eqnarray}
(\pi ^{(d-1)}\otimes 1)L_{13}^{\ast }(zq^{d})R_{23}^{\ast }(z)(\imath
\otimes 1) &=&(z^{-1}q-q^{-1})(\imath \otimes 1)(\pi ^{(d-2)}\otimes
1)L^{\ast }(zq^{d+1})  \notag \\
(p\otimes 1)(\pi ^{(d-1)}\otimes 1)L_{13}^{\ast }(zq^{d})R_{23}(z)
&=&(z^{-1}-1)(\pi ^{(d)}\otimes 1)L^{\ast }(zq^{d-1})(p\otimes 1)
\end{eqnarray}%
where the maps%
\begin{equation}
\imath :\;\left\vert k\right\rangle \hookrightarrow \lbrack
d-k-1]_{q}\left\vert k\right\rangle \otimes \left\vert 1\right\rangle
-q^{d-k-1}[k+1]_{q}\left\vert k+1\right\rangle \otimes \left\vert
0\right\rangle 
\end{equation}%
and%
\begin{equation}
p:\;\frac{[d]}{[d-k]}~\left\vert k\right\rangle \otimes \left\vert
0\right\rangle \rightarrow \left\vert k\right\rangle 
\end{equation}%
are the inclusion and projection map in the exact sequence (\ref{seq1}).

\section{The Quantum Q-operator}

After outlining the derivation of (\ref{quantumFus}) we now turn to the $Q$%
-operator and apply the same strategy as in the case of the fusion
hierarchy. The main difference lies in the fact that for the definition of
the Q-operator we need to introduce an \emph{infinite}-dimensional
evaluation module when $q$ is generic (i.e. not a root of unity) \cite{RW02} 
\cite{CKQ3},%
\begin{eqnarray}
\rho ^{+}(e_{0})\left\vert k\right\rangle &=&z\left\vert k+1\right\rangle
,\quad \rho ^{+}(q^{\frac{h_{1}}{2}})\left\vert k\right\rangle =\rho
^{+}(q^{-\frac{h_{0}}{2}})\left\vert k\right\rangle =r^{\frac{1}{2}%
}q^{-k-1/2}\left\vert k\right\rangle ,  \notag \\
\rho ^{+}(e_{1})\left\vert k\right\rangle &=&\frac{s+1-q^{2k}-sq^{-2k}}{%
(q-q^{-1})^{2}}\;\left\vert k-1\right\rangle ,\quad \rho
^{+}(e_{1})\left\vert 0\right\rangle =0,\quad r,s,z\in \mathbb{C}~.
\label{rho}
\end{eqnarray}%
Here $s,r$ are free parameters characterizing the representation and $z$
denotes the spectral variable as before. In connection with spin-reversal
one also encounters the module 
\begin{equation}
\rho ^{-}:=\rho ^{+}\circ \omega \quad \text{with\quad }\{e_{1},e_{0},q^{%
\frac{h_{1}}{2}},q^{\frac{h_{0}}{2}}\}\overset{\omega }{\rightarrow }%
\{e_{0},e_{1},q^{\frac{h_{0}}{2}},q^{\frac{h_{1}}{2}}\}\;.  \label{rhominus}
\end{equation}%
Note that in the limit $s\rightarrow 0$ we recover the $q$-oscillator
representations used in \cite{BLZ97}. In the case that $q$ is a primitive
root of unity of order $\ell $ we truncate the evaluation module $\rho ^{+}$
by imposing the condition (compare with \cite{CKQ4})%
\begin{equation}
q^{\ell }=1:\qquad \rho ^{+}(e_{0})\left\vert \ell ^{\prime }-1\right\rangle
=0,\quad \ell ^{\prime }=\left\{ 
\begin{array}{cc}
\ell , & \text{if }\ell \text{ is odd} \\ 
\ell /2, & \text{if }\ell \text{ is even}%
\end{array}%
\right. \;.  \label{trunc}
\end{equation}%
The intertwiner corresponding to the quantum group module $\rho ^{+}\otimes
\pi $ has been computed previously \cite{RW02} \cite{CKQ3} and reads,%
\begin{equation}
\mathfrak{L}(z)=\left( 
\begin{array}{cc}
z\frac{s}{r}~q^{\frac{h_{1}+1}{2}}-q^{-\frac{h_{1}+1}{2}} & (q-q^{-1})q^{%
\frac{h_{1}+1}{2}}e_{0} \\ 
(q-q^{-1})e_{1}q^{-\frac{h_{1}+1}{2}} & zr~q^{-\frac{h_{1}-1}{2}}-q^{\frac{%
h_{1}-1}{2}}%
\end{array}%
\right) \in U_{q}(\widehat{sl}_{2})\otimes \limfunc{End}V\ .  \label{LQ0}
\end{equation}%
In order to define a $Q$-operator for the quantum transfer matrix we now
need to compute the intertwiner corresponding to the module $\rho
^{+}\otimes \pi ^{\ast }$. The result is%
\begin{equation}
\mathfrak{L}^{\ast }(z)=\left( 
\begin{array}{cc}
zr~q^{-\frac{h_{1}+1}{2}}-q^{\frac{h_{1}+1}{2}} & (q^{-1}-q)e_{1}q^{-\frac{%
h_{1}+1}{2}} \\ 
(q^{-1}-q)q^{\frac{h_{1}+1}{2}}e_{0} & z\frac{s}{r}~q^{\frac{h_{1}-1}{2}%
}-q^{-\frac{h_{1}-1}{2}}%
\end{array}%
\right) \in U_{q}(\widehat{sl}_{2})\otimes \limfunc{End}V^{\ast }\ .
\label{LQ}
\end{equation}%
The last expression is derived from the coproduct relations%
\begin{eqnarray}
\Delta ^{\ast }(e_{1}) &=&e_{1}\otimes 1-q^{1+h_{1}}\otimes \sigma
^{-},\quad \quad \Delta _{\text{op}}^{\ast }(e_{1})=e_{1}\otimes q^{-\sigma
^{z}}-q~1\otimes \sigma ^{-},  \notag \\
\Delta ^{\ast }(e_{0}) &=&e_{0}\otimes 1-q^{1+h_{0}}\otimes \sigma
^{+},\quad \quad \Delta _{\text{op}}^{\ast }(e_{0})=e_{0}\otimes q^{\sigma
^{z}}-q~1\otimes \sigma ^{+}\ .
\end{eqnarray}%
We define for $r=1$ in $\rho ^{+}$ the operator%
\begin{equation}
Q(z;s)=\limfunc{Tr}_{\rho ^{+}}q^{\alpha h_{1}\otimes 1}\mathfrak{L}_{N}(zw)%
\mathfrak{L}_{N-1}^{\ast }(z/w)\cdots \mathfrak{L}_{2}(zw)\mathfrak{L}%
_{1}^{\ast }(z/w),\quad q^{\alpha }=e^{\beta h/2},  \label{Q}
\end{equation}%
where we have normalised (\ref{LQ}) such that $Q$ is polynomial in the
spectral parameter $z$. The specialization to $r=1$ can be imposed without
any loss loss of generality due to the identity%
\begin{equation}
Q(z;r,s)=r^{\alpha -S_{A}}Q(z;r=1,s)\;.
\end{equation}%
Having stated the explicit definition of the $Q$-operator we now turn to its
properties and the functional relations it satisfies. Since the auxiliary
spaces of the quantum transfer matrix and the $Q$-operator are the same as
in the conventional, classical six-vertex model the proofs for the
statements made below follow closely the line of argument presented
previously in \cite{CKQ3,CKQ4,CKQ7}. I shall therefore omit the proofs from
the main text and refer the reader to the appendix for further details.

\subsection{Factorisation}

One important result is that the $Q$-operator factorises into simpler
operators as follows (we still assume $\alpha \neq 0$),%
\begin{equation}
Q(z;s)=Q(0;s)Q^{+}(z)Q^{-}(zs),  \label{Qfactor}
\end{equation}%
where 
\begin{equation}
Q^{+}(z)=\lim_{s\rightarrow 0}Q(0;s)^{-1}Q(z;s)  \label{Qplus}
\end{equation}%
and the normalisation factor $Q(0;s)$ at $z=0$ is easily computed to be%
\begin{equation}
\lim_{z\rightarrow 0}Q(z;s)=\limfunc{Tr}_{\rho ^{+}}q^{(\alpha
-S_{A})h_{1}}=\left\{ 
\begin{array}{cc}
\dfrac{1}{q^{\alpha -S_{A}}-q^{S_{A}-\alpha }},\medskip & \;\;q\;\text{%
generic} \\ 
\dfrac{1-q^{2\ell (S_{A}-\alpha )}}{q^{\alpha -S_{A}}-q^{S_{A}-\alpha }}, & 
\;\;q^{\ell }=\pm 1%
\end{array}%
\right. \ .  \label{norm}
\end{equation}%
The identity (\ref{norm}) for generic $q$ has to be understood as analytic
continuation from the region of convergence. Note that the point $\alpha =0$%
, i.e. vanishing magnetic field $h=0$, remains singular. This can be
understood from the construction of the $Q$-operator since for generic $q$
the auxiliary space given by the the representation (\ref{rho}) is
infinite-dimensional and twisted boundary conditions with an appropriate
choice of $\alpha $ are needed to ensure that the trace in (\ref{Q}) is
well-defined; compare with the discussion in \cite{CKQ3}. The remaining
operator $Q^{-}$ is the counterpart of $Q^{+}$ and can be obtained by
employing the following transformation,%
\begin{equation}
Q(z^{-1},w^{-1};s)=z^{-N}s^{\frac{N}{2}+S_{A}}~(w/q)^{-S^{z}}\mathfrak{R}%
~Q(z/s,wq^{-2};s)^{t}~\mathfrak{R}(w/q)^{S^{z}}\;.  \label{RQR}
\end{equation}%
Here the operators $S^{z},\mathfrak{R}$ have been introduced earlier in (\ref%
{RS}) and use has been made of the identities%
\begin{eqnarray*}
\mathfrak{L}(z^{-1}) &=&-z^{-1}q(1\otimes (-\tfrac{z}{sq})^{-\sigma
^{z}}\sigma ^{x})\mathfrak{L}(zq^{-2}/s)^{1\otimes t}(1\otimes \sigma ^{x}(-%
\tfrac{z}{sq})^{\sigma ^{z}})s^{\frac{1+\sigma ^{z}}{2}} \\
\mathfrak{L}^{\ast }(z^{-1}) &=&-z^{-1}q^{-1}(1\otimes (-\tfrac{zq}{s}%
)^{\sigma ^{z}}\sigma ^{x})\mathfrak{L}^{\ast }(zq^{2}/s)^{1\otimes
t}(1\otimes \sigma ^{x}(-\tfrac{zq}{s})^{-\sigma ^{z}})s^{\frac{1-\sigma ^{z}%
}{2}}\;.
\end{eqnarray*}%
The identity (\ref{RQR}) then implies that up to an unimportant
normalization constant we have,%
\begin{equation}
Q^{-}(z,w)\propto z^{\frac{N}{2}+S_{A}}~(wq)^{S^{Z}}\mathfrak{R}%
~Q^{+}(z^{-1},w^{-1}q^{-2})^{t}~\mathfrak{R}(wq)^{-S^{Z}}~.
\end{equation}%
Note that the action of the spin-reversal operator relates $Q^{-}$ to the
representation (\ref{rhominus}). Both operators $Q^{\pm }$ have polynomial
eigenvalues w.r.t. the spectral variable $z$. I shall denote these
eigenvalues by the same symbol as the operators,%
\begin{equation}
Q^{\pm }(z)=\prod_{k=1}^{n_{\pm }}(1-x_{k}^{\pm }z)=\sum_{k=0}^{n_{\pm
}}e_{k}^{\pm }(-z)^{k},\qquad n_{\pm }=n\mp S_{A}\;.  \label{Qpm}
\end{equation}%
As we will see below the polynomial roots $x_{k}^{\pm }$ are two sets of
Bethe roots. They can be (numerically) computed by employing a number of
functional relations which are satisfied by the $Q$-operator.

\subsection{Functional relations}

The best known functional relation is the generalization of Baxter's $TQ$%
-equation for the six-vertex model. This equation is obtained in the present
construction for the $Q$-operator (which differs from Baxter's approach) by
first deriving the functional relation%
\begin{equation}
Q(z;s)\tau ^{(1)}(z)=q^{\alpha -S_{A}}\tau
^{(0)}(zq)Q(zq^{-2};sq^{2})+q^{S_{A}-\alpha }\tau
^{(0)}(zq^{-1})Q(zq^{2};sq^{-2})
\end{equation}%
which is a direct consequence of the following decomposition of the tensor
product of representations, $\rho ^{+}(z;r,s)\otimes \pi _{z},$ described by
the exact sequence \cite{RW02} \cite{CKQ3}%
\begin{equation}
0\rightarrow \rho ^{+}(zq^{2};rq^{-1},sq^{-2})\hookrightarrow \rho
^{+}(z;r,s)\otimes \pi _{z}\rightarrow \rho
^{+}(zq^{-2},rq,sq^{2})\rightarrow 0\ .
\end{equation}%
Here $\tau ^{(0)}$ is the quantum determinant introduced in (\ref{quantumdet}%
). Taking the limit $s\rightarrow 0$ employing (\ref{Qplus}) we obtain the $%
TQ$-equation,%
\begin{equation}
Q^{+}(z)\tau ^{(1)}(z)=q^{\alpha -S_{A}}\tau
^{(0)}(zq)Q^{+}(zq^{-2})+q^{S_{A}-\alpha }\tau ^{(0)}(zq^{-1})Q^{+}(zq^{2})\
.  \label{TQplus}
\end{equation}%
A similar relation holds for $Q^{-}$ when employing the transformations (\ref%
{RtauR}), (\ref{tautransp}) and (\ref{RQR}). The $TQ$ equation holds true
also for vanishing external magnetic field, i.e. $\alpha =0$. If the
magnetic field is nonzero, however, there is another identity which makes
use of both solutions $Q^{\pm }$ and on which we will focus. It yields a
simple expression for all elements in the fusion hierarchy in terms of $%
Q^{\pm }$,%
\begin{equation}
\tau ^{(d-1)}(z)=\frac{q^{d(\alpha
-S_{A})}Q^{+}(zq^{-d})Q^{-}(zq^{d})-q^{d(S_{A}-\alpha
)}Q^{+}(zq^{d})Q^{-}(zq^{-d})}{q^{\alpha -S_{A}}-q^{S_{A}-\alpha }}\;.
\label{Fusion&Q}
\end{equation}%
The last expression can even be analytically continued to complex dimension $%
d$ if one wishes to make contact with the trace functional introduced in the
context of correlation functions for the infinite XXZ spin-chain. Here we
shall not pursue this aspect further but refer the reader to \cite{BJMST}\
and references therein; see also \cite{CKQ7}.

\section{The Wronskian relation}

Note that when setting $d=1$ in (\ref{Fusion&Q}) the left hand side of the
above equation is explicitly known and we arrive at%
\begin{equation}
(zwq-1)^{n}(z/wq-1)^{n}=\frac{q^{\alpha
-S_{A}}Q^{+}(zq^{-1})Q^{-}(zq)-q^{S_{A}-\alpha }Q^{+}(zq)Q^{-}(zq^{-1})}{%
q^{\alpha -S_{A}}-q^{S_{A}-\alpha }}\ .  \label{quantumW0}
\end{equation}%
This functional relation, which is the discrete analogue of a Wronskian in
the theory of second order differential equations, can therefore be employed
to compute the eigenvalues of $Q^{\pm }$. It is this observation which we
will investigate further in the remainder of this paper.

\subsection{System of quadratic equations}

Expanding the Wronskian relation (\ref{quantumW0}) with respect to the
spectral parameter $z$ yields the following system of $N$-quadratic
equations for the unknown coefficients $e_{k}^{\pm }$ which are the
elementary symmetric polynomials in the Bethe roots,%
\begin{equation}
\sum_{k+l=m}\binom{~n~}{k}\binom{~n~}{l}(wq)^{k-l}=\sum_{k+l=m}\frac{%
q^{\alpha -S_{A}-k+l}-q^{S_{A}-\alpha +k-l}}{q^{\alpha
-S_{A}}-q^{S_{A}-\alpha }}~e_{k}^{+}e_{l}^{-}\ .  \label{quantumW}
\end{equation}%
It should be emphasized once more that this set of equations contains for $%
\alpha \neq 0$ all the necessary information about the spectrum of the
quantum transfer matrix as well as the higher spin transfer matrices. Recall
that $N$ is the Trotter number, $w=e^{-\beta ^{\prime }/N}=\exp (-\frac{%
\beta (q-q^{-1})}{N})$ contains the temperature variable and $q^{\alpha
}=e^{h\beta /2}$ the external magnetic field. Setting now $n=2$ in (\ref%
{Fusion&Q}) the eigenvalues of the quantum transfer matrix (\ref{quantumT})
are obtained from the identity 
\begin{equation}
\tau (z;w)=\frac{q^{2(\alpha
-S_{A})}Q^{+}(zq^{-2})Q^{-}(zq^{2})-q^{2(S_{A}-\alpha
)}Q^{+}(zq^{2})Q^{-}(zq^{-2})}{(q^{\alpha -S_{A}}-q^{S_{A}-\alpha
})(zwq-q^{-1})^{n}(z/wq-q)^{n}}\ .  \label{TQpm}
\end{equation}%
The equations (\ref{quantumW}) and (\ref{TQpm}) are the aforementioned
generalisations of the identities (\ref{result}) and (\ref{LamN}) in the
introduction.

\subsubsection{The Bethe ansatz equations}

Having identified (\ref{quantumW}) as the key relation in describing the
spectrum of the quantum transfer matrix for finite $N$ we need to discuss
the relation with the Bethe ansatz equations which are usually considered to
be the fundamental set of identities for discussing the spectrum. Starting
from the quantum Wronskian we set 
\begin{eqnarray*}
z &=&q/x_{i}^{+}:\qquad \frac{-q^{S_{A}-\alpha
}Q^{+}(q^{2}/x_{i}^{+})Q^{-}(1/x_{i}^{+})}{q^{\alpha -S_{A}}-q^{S_{A}-\alpha
}}=(wq^{2}/x_{i}^{+}-1)^{n}(1/wx_{i}^{+}-1)^{n} \\
z &=&q^{-1}/x_{i}^{+}:\qquad \frac{q^{\alpha
-S_{A}}Q^{+}(q^{-2}/x_{i}^{+})Q^{-}(1/x_{i}^{+})}{q^{\alpha
-S_{A}}-q^{S_{A}-\alpha }}=(w/x_{i}^{+}-1)^{n}(q^{-2}/wx_{i}^{+}-1)^{n}
\end{eqnarray*}%
and obtain%
\begin{equation}
-q^{-2\alpha }\prod_{j=1}^{n_{+}}\frac{x_{j}^{+}q/x_{i}^{+}-q^{-1}}{%
x_{j}^{+}q^{-1}/x_{i}^{+}-q}=\left( \frac{wq-x_{i}^{+}q^{-1}}{%
q^{-1}-wx_{i}^{+}q}\right) ^{n}\left( \frac{1-wx_{i}^{+}}{w-x_{i}^{+}}%
\right) ^{n}\ .  \label{algBAE}
\end{equation}%
The last equation is one among the $n_{+}$ Bethe ansatz equations as they
can be found for instance in \cite{Klue04}; see equations (29) and (30). To
facilitate the comparison note that under the parametrisation%
\begin{equation*}
x_{i}=e^{\gamma \lambda _{i}},\qquad q=e^{i\gamma },\qquad w=e^{-i\gamma
\tau }=e^{-(q-q^{-1})\beta /N},\qquad q^{2\alpha }=e^{\beta h}
\end{equation*}%
the Bethe ansatz equations are rewritten as%
\begin{multline*}
\frac{\phi (\lambda _{j}+i)}{\phi (\lambda _{j}-i)}:=\frac{\sinh ^{\frac{N}{2%
}}\frac{\gamma }{2}(\lambda _{j}-i\tau +2i)\sinh ^{\frac{N}{2}}\frac{\gamma 
}{2}(\lambda _{j}+i\tau )}{\sinh ^{\frac{N}{2}}\frac{\gamma }{2}(\lambda
_{j}+i\tau -2i)\sinh ^{\frac{N}{2}}\frac{\gamma }{2}(\lambda _{j}-i\tau )}=
\\
-e^{\beta h}\prod_{k=1}^{n_{+}}\frac{\sinh \frac{\gamma }{2}(\lambda
_{j}-\lambda _{k}+2i)}{\sinh \frac{\gamma }{2}(\lambda _{j}-\lambda _{k}-2i)}%
=:-e^{\beta h}\frac{\mathcal{Q}(\lambda _{j}+2i)}{\mathcal{Q}(\lambda
_{j}-2i)}
\end{multline*}%
which is the notation used in \cite{Klue04}. From this we infer that the
Wronskian relation (\ref{quantumW0}) implies the Bethe ansatz equations, the
converse is not true. Note that $Q^{-}$ yields another set of Bethe roots,
which in terms of the algebraic Bethe ansatz (see the appendix) allow one to
construct the eigenvector from the lowest (instead of the highest) weight
vector.

\subsection{Special solutions in the $S_{A}=0$ sector}

As pointed out earlier the quantum transfer matrix block decomposes with
respect to the alternating spin-operator $S_{A}$. In particular, the largest
eigenvalue $\Lambda _{N}$ appears to be in the $S_{A}=0$ sector. Since $%
\lim_{N\rightarrow \infty }\Lambda _{N}$ yields the partition function (\ref%
{Z}) in the thermodynamic limit its corresponding solutions to (\ref%
{quantumW}) are of particular, physical interest.

Quite generally, we infer from (\ref{quantumW0}) that there are certain
symmetries among the solutions to the Wronskian relation, for instance when
replacing $z\rightarrow z^{-1}$. It is then natural to assume that some
solutions are invariant under these transformations, especially when they
belong to non-degenerate or distinguished eigenvalues such as $\Lambda _{N}$%
. Thus, one would expect that there is a subset of solutions for which%
\begin{equation}
Q^{\pm }(z)=\prod_{i=1}^{n}(1-z/x_{i}^{\mp })  \label{recid}
\end{equation}%
holds. That is, $Q^{\pm }$ are the inverse or reciprocal polynomial of each
other. In fact, one verifies numerically for Trotter numbers up to 16 that
in the $S_{A}=0$ sector there are always $2^{n}$ such "special" solutions
(for $q$ real or on the unit circle) and that among this set of solutions is
the one describing the largest eigenvalue of the quantum transfer matrix.
Conjecturing this to be true for all $N$ one can then halve the number of
variables in (\ref{quantumW}) arriving at the result (\ref{result}), (\ref%
{LamN}) stated in the introduction. A similar observation has been made
previously for the twisted XXX spin-chain; see Section 5.1 in \cite{CKQXXX}.

Provided that $q$ (and therefore $w=\exp [(q-q^{-1})\beta /N]$) lies on the
unit circle there is another obvious "symmetry" of (\ref{quantumW0}),
complex conjugation. In fact, one finds for $S_{A}=0$ that there exist
solutions $Q^{\pm }$ invariant under this transformation, i.e. obeying the
additional restriction%
\begin{equation}
Q^{\pm }(z)=\prod_{i=1}^{n}(1-(x_{i}^{\mp })^{\ast }z),\quad z\in \mathbb{R}%
\;.  \label{ccid}
\end{equation}%
In terms of the elementary symmetric polynomials $e_{k}^{\pm }$ the
restrictions (\ref{recid}) and (\ref{ccid}) correspond to the identities%
\begin{equation}
e_{k}^{\pm }=e_{n-k}^{\mp }/e_{n}^{\mp }\qquad \text{and}\qquad e_{k}^{\pm
}=(e_{k}^{\mp })^{\ast },
\end{equation}%
respectively. Whether these solutions persist for Trotter numbers $N>16$
needs to be investigated numerically. We leave this to future work.

\section{Conclusions and outlook}

In this paper Baxter's $Q$-operator has been constructed for the quantum
transfer matrix of the XXZ spin-chain. The main motivation has been to
derive a set of equations which allow one to describe the spectrum of the
quantum transfer matrix and are simpler than the previously known Bethe
ansatz equations. For non-vanishing external magnetic field this is indeed
possible and the order of the equations can be reduced from $N$ (Bethe
ansatz equations (\ref{algBAE})) to order two (the Wronskian relation (\ref%
{quantumW})) which is a great simplification for numerical computations. Of
particular interest is the largest eigenvalue of the quantum transfer matrix
which in the thermodynamic limit (i.e. the physical system size $L$ tends to
infinity) contains all the relevant finite temperature behaviour of the
spin-chain. For this eigenvalue we observed further simplifications in its
polynomial structure which lead to a reduction of the number of variables by
a factor two. The latter observation is based on symmetries of the Wronskian
relation (\ref{quantumW0}) and numerical computations which were carried out
up to the Trotter number $N=16$. To go beyond this bound requires more
extensive numerical computations which are planned to be carried out in
future work.

Another aspect which has been omitted from the present work is the
derivation of integral equations similar to those non-linear integral
equations which have been previously obtained on the basis of the Bethe
ansatz equations; see e.g. \cite{Klue04} and references therein. The basis
for such a derivation is numerical evidence for the distribution of the
Bethe roots in the large Trotter number limit, $N\rightarrow \infty $. For
this reason the discussion of integral equations and their connection with
the thermodynamic Bethe ansatz are postponed until the necessary numerical
data has been obtained.

For vanishing magnetic field the Wronskian relation ceases to be valid but
in the text it was shown that at roots of unity one might be able to employ
the same loop algebra symmetry as the one which exists for the six-vertex
model. This might possibly help to reduce the order of equations or simplify
the computation of the spectrum of the quantum transfer matrix.\medskip 
\newline

\noindent \textbf{Acknowledgments: }The author wishes to thank Robert Weston
for comments on a draft version of this paper and the organisers of the
EUCLID 2006 conference, September 11-15, ENS Lyon, France, where some of the
results in this paper have been presented. This work has been financially
supported by a University Research Fellowship of the Royal Society.

\bibliographystyle{phreport}
\bibliography{RefQ}

\appendix

\section{Properties of $Q$ for generic $q$}

For comparison with the line of argument for the classical six-vertex model
the reader might wish to consult \cite{CKQ3}. The line of argument closely
follows the exposition given there.

\subsection{The fusion hierarchy in terms of $Q$}

Setting the free parameter $s$ in $\rho ^{+}$ to the special value $%
s=q^{2d},\;d=1,2,...$ one derives the following identities by restricting
the $\mathfrak{L}$ and $\mathfrak{L}^{\ast }$-operator given in (\ref{LQ0})
and (\ref{LQ}) to the subspaces $V_{<d}=\limfunc{span}\{\left\vert
k\right\rangle \}_{k=0}^{d-1}$ and $V_{\geq d}=\limfunc{span}\{\left\vert
k\right\rangle \}_{k=d}^{\infty }$,%
\begin{eqnarray}
s &=&q^{2d}:\;\mathfrak{L}^{\ast }(z)|_{V_{<d}}=q^{d}(1\otimes q^{\frac{d}{2}%
\sigma ^{z}})L^{\ast (d-1)}(zq^{d})(1\otimes q^{-d\sigma ^{z}}) \\
s &=&q^{2d}:\;\mathfrak{L}^{\ast }(z)|_{V_{\geq d}}=q^{2d}(1\otimes
q^{d\sigma ^{z}})\mathfrak{L}^{\ast }(zq^{2d};s\rightarrow q^{-2d})(1\otimes
q^{-2d\sigma ^{z}})
\end{eqnarray}%
Employing that by construction the $Q$-operator commutes with the
alternating spin-operator, $[Q,S_{A}]=0$, one deduces from these last two
equations the following identity for the higher spin quantum transfer
matrices,%
\begin{equation}
\tau ^{(d-1)}(z)=q^{d(\alpha -S_{A})}Q(zq^{-d};s=q^{2d})-q^{d(S_{A}-\alpha
)}Q(zq^{d};s=q^{-2d})\;.
\end{equation}%
Below we will see that this relation can be simplified further due to a
factorisation of the $Q$-operator into two parts.

\subsection{Algebraic Bethe ansatz computation}

In order to obtain the eigenvalues we apply the algebraic Bethe ansatz for
the quantum transfer matrix (see e.g. \cite{GKS04}) and compute the action
of $Q$ on a Bethe state. Denote by $\{\uparrow ,\downarrow \}$ by the
orthogonal basis in $\mathbb{C}^{2}$ and consider the reference state 
\begin{equation}
\left\vert 0\right\rangle =\downarrow \otimes \uparrow \cdots \otimes
\downarrow \otimes \uparrow \ .  \label{refstate}
\end{equation}%
Let $\{A,B,C,D\}$ be the matrix elements of the quantum monodromy matrix (%
\ref{mom}) then a Bethe state is given by%
\begin{equation*}
\left\vert x_{1}^{+},...,x_{n_{+}}^{+}\right\rangle :=B(1/x_{1}^{+})\cdots
B(1/x_{n_{+}}^{+})\left\vert 0\right\rangle
\end{equation*}%
with $\{x_{1}^{+},...,x_{n_{+}}^{+}\}$ being a solution to the Bethe ansatz
equations (\ref{algBAE}). Since the auxiliary space of the quantum transfer
matrix (\ref{quantumT}) and $Q$-operator (\ref{Q}) are the same as in the
ordinary case of the classical six-vertex transfer matrix with
quasi-periodic boundary conditions the results from \cite{CKQ3} apply. Using
the commutation relations between the matrix elements $A,B,C,D$ and those of
the monodromy matrix of the $Q$-operator, 
\begin{equation}
\boldsymbol{Q}_{k,l}(z;s):=\left\langle k\right\vert q^{\alpha h_{1}\otimes
1}\mathfrak{L}_{N}(zw)\mathfrak{L}_{N-1}^{\ast }(z/w)\cdots \mathfrak{L}%
_{2}(zw)\mathfrak{L}_{1}^{\ast }(z/w)\left\vert l\right\rangle ,
\label{Qmom}
\end{equation}%
detailed in \cite{CKQ3} we obtain%
\begin{multline*}
Q(z;s)\left\vert x_{1}^{+},...,x_{n_{+}}^{+}\right\rangle = \\
\left\{ \sum_{k=0}^{\infty }\left\langle 0\right\vert \boldsymbol{Q}%
_{kk}(z;s)\left\vert 0\right\rangle \tprod_{j=1}^{n_{+}}\frac{\left\langle
k+1\right\vert \mathfrak{a}_{j}\left\vert k+1\right\rangle \left\langle
k\right\vert \mathfrak{d}_{j}\left\vert k\right\rangle -\left\langle
k\right\vert \mathfrak{c}_{j}\mathfrak{b}_{j}\left\vert k\right\rangle }{%
\left\langle k+1\right\vert \mathfrak{a}_{j}\left\vert k+1\right\rangle
\left\langle k\right\vert \mathfrak{a}_{j}\left\vert k\right\rangle }%
\right\} ~\left\vert x_{1}^{+},...,x_{n_{+}}^{+}\right\rangle
\end{multline*}%
Here we have introduce the abbreviations%
\begin{equation}
\mathfrak{L}(zx_{j}^{+})=\left( 
\begin{array}{cc}
\mathfrak{a}_{j} & \mathfrak{b}_{j} \\ 
\mathfrak{c}_{j} & \mathfrak{d}_{j}%
\end{array}%
\right)  \label{Labcd}
\end{equation}%
for the matrix elements of the $\mathfrak{L}$-operator. Inserting the
explicit expressions for the latter which can be read off from (\ref{LQ0})
one arrives at%
\begin{multline*}
Q(z;s)\left\vert x_{1}^{+},...,x_{n_{+}}^{+}\right\rangle = \\
\left\{ \sum_{k=0}^{\infty }\left\langle 0\right\vert \boldsymbol{Q}%
_{kk}(z;s)\left\vert 0\right\rangle \tprod_{j=1}^{n_{+}}\frac{%
q^{-2k-1}(1-z~x_{j}^{+})(1-zs~x_{j}^{+})}{%
(1-zsq^{-2k}~x_{j}^{+})(1-zsq^{-2k-2}~x_{j}^{+})}\right\} ~\left\vert
x_{1}^{+},...,x_{n_{+}}^{+}\right\rangle \\
=q^{S_{A}-\alpha }Q^{+}(z)Q^{+}(zs)\sum_{k=0}^{\infty }\frac{%
q^{2k(S_{A}-\alpha )}\tau ^{(0)}(zsq^{-2k-1})}{%
Q^{+}(zsq^{-2k})Q^{+}(zsq^{-2k-2})}~\left\vert
x_{1}^{+},...,x_{n_{+}}^{+}\right\rangle \ .
\end{multline*}%
In the last line we have used the definition of $Q^{+}$ as a polynomial (\ref%
{Qpm}) and%
\begin{equation*}
\left\langle 0\right\vert \boldsymbol{Q}_{kk}(z;s)\left\vert 0\right\rangle
=q^{(n-\alpha )(2k+1)}[(zwsq^{-2k}-1)(zsq^{-2k-2}/w-1)]^{n}\ .
\end{equation*}%
Taking the limit $s\rightarrow 0$ in order to fix the normalisation constant
we deduce from this expression the following formula for $Q^{-}$,%
\begin{equation}
Q^{-}(z)=Q^{+}(z)\sum_{k=0}^{\infty }\frac{q^{2k(S_{A}-\alpha )}\tau
^{(0)}(zq^{-2k-1})}{Q^{+}(zq^{-2k})Q^{+}(zq^{-2k-2})}\ .  \label{ABAQminus}
\end{equation}%
Note that in \cite{CKQ3} the vanishing of the unwanted terms in the Bethe
ansatz has only been verified for $n_{+}=1,2,3$, since the algebraic Bethe
ansatz computation is more involved than in the case of the transfer matrix
due to the infinite-dimensional auxiliary space of the $Q$-operator.
However, the result for the spectrum coincides with the one found at roots
of unity and the eigenvalues satisfy all aforementioned functional relations
for the $Q$-operator, which have been derived by different means. We shall
take this as sufficient evidence that the algebraic Bethe ansatz computation
presented above holds also true for $n_{+}>3$.

\section{Properties of $Q$ when $q$ is a root of unity}

Let us now turn to the case when $q$ is a primitive root of unity of order $%
\ell $.

\subsection{Functional relations for the fusion hierarchy}

The evaluation module $\rho ^{+}$ is now finite-dimensional according to (%
\ref{trunc}). For analysing the spectrum of $Q$ we now rely on a functional
relation derived in \cite{CKQ4}. To connect with the discussion therein we
set $r=\mu ^{-1}$ and $s=\mu ^{-2}$ in (\ref{rho}) and obtain the evaluation
representation $\pi _{z}^{\mu }$ specified in equation (15-16) of \cite{CKQ4}%
. Thus, in the following we refer to $\rho ^{+}(r=\mu ^{-1},s=\mu ^{-2})$ as 
$\pi _{z}^{\mu }$. Then the following short exact sequence holds (see
equations (52-53) in \cite{CKQ4}), 
\begin{equation}
0\rightarrow \pi _{\mu q}^{\mu \nu q}\rightarrow \pi _{\mu \nu q^{2}}^{\mu
}\otimes \pi _{1}^{\nu }\rightarrow \pi _{\mu q^{-\ell ^{\prime }+1}}^{\mu
\nu q^{-\ell ^{\prime }+1}}\otimes \pi _{\nu q^{\ell ^{\prime }+1}}^{(\ell
^{\prime }-2)}\rightarrow 0\ .
\end{equation}%
Here $\pi _{z}^{(\ell ^{\prime }-2)}$ is the evaluation representation of
spin $(\ell ^{\prime }-1)/2$. As before this decomposition of the tensor
product of representations implies a functional relation, but this time it
involves a product of two Q-operators with different values for the free
parameters entering (\ref{rho}). Setting $s=\mu ^{-2}$ and $t=\nu ^{-2}$
this functional relation reads%
\begin{equation}
Q(zq^{2}/s;s)Q(z;t)=q^{S_{A}-\alpha
}Q(zq^{2}/s;stq^{-2})[(zq^{2}-1)^{n}(z-1)^{n}+q^{\ell ^{\prime
}(S_{A}-\alpha )}\tau ^{(\ell ^{\prime }-2)}(zq^{\ell ^{\prime }+1})]\ ,
\label{QQfunc}
\end{equation}%
compare with equation (46) in \cite{CKQ7}. Assume that $%
[Q(z_{1};s_{1}),Q(z_{2};s_{2})]=0$ for arbitrary pairs $z_{1,2},s_{1,2}\in 
\mathbb{C}$. This has been proved for $\ell =3,4,6$ by explicitly
constructing the corresponding intertwiner for $\rho
^{+}(z_{1},s_{1})\otimes \rho ^{+}(z_{2};s_{2})$ \cite{CKQ2,CKQProc}. Thus,
the eigenvalues of $Q(z;s)$ must be polynomial in $z$ (and $s$) their most
general form being%
\begin{equation*}
Q(z;s)=\frac{1-q^{2\ell ^{\prime }(S_{A}-\alpha )}}{q^{\alpha
-S_{A}}-q^{S_{A}-\alpha }}\tprod_{j=1}^{k}(1-z~x_{j})%
\tprod_{j=1}^{N-k}(1-z~y_{j}(s))\ .
\end{equation*}%
Here we have taken the limit $z\rightarrow 0$ to fix the normalisation
constant,%
\begin{equation}
\lim_{z\rightarrow 0}Q(z;s)=\limfunc{Tr}_{\rho ^{+}}q^{(\alpha
-S_{A})h_{1}}\ .
\end{equation}%
We assume that the roots $x_{j}=x^+_{j}$ are independent of $s$ while the $y_{j}$'s
depend on it allowing for the possibilities that either $k=0$ or $k=N$.
Inserting this general expression for the eigenvalue into (\ref{QQfunc}) one
deduces that the roots $y_{j}(s)$ can only depend linearly on $s$, i.e. $%
y_{j}(s)=x_{j}^{-}s$ for some $x_{j}^{-}$. This implies the factorisation (%
\ref{Qfactor}) in the text. Futhermore we infer from (\ref{RQR}) that $%
k=N/2-S_{A}$ in each fixed $S_{A}$ sector.

\end{document}